\documentstyle[prl,aps,multicol]{revtex}
\begin{document}
\bibliographystyle{prsty}
\title{First Order Premelting Transition of Vortex
Lattices}
\author{Herv\'e M. Carruzzo and Clare C. Yu}
\address{
Department of Physics and Astronomy, University of California,
Irvine, Irvine, California 92697}
\date{\today}
\maketitle
\begin{abstract}
Vortex lattices in the high temperature
superconductors undergo a first order phase transition which
has thus far been regarded as melting 
from a solid to a liquid. We point out an alternative possibility
of a two step process in which there is a first order 
transition from an ordinary vortex lattice to a soft vortex solid
followed by another first order melting transition from the
soft vortex solid to a vortex liquid. We focus on the first
step. This premelting transition is induced 
by vacancy and interstitial vortex 
lines. We obtain good agreement with the experimental transition
temperature versus field, latent heat, and 
magnetization jumps for YBCO and BSCCO. 
\end{abstract}


\begin{multicols}{2}
\narrowtext
Phase transitions involving
vortex lattices in the high temperature superconductors
is an area of active study 
\cite{review:blatter,review:brandt}.
Below a critical value of the magnetic field, 
vortex lattices in
YBa$_{2}$Cu$_{3}$O$_{7-\delta}$ (YBCO) 
\cite{safar93,liang96,welp96,schilling96} and 
Bi$_{2}$Sr$_{2}$CaCu$_{2}$O$_{8}$ (BSCCO) 
\cite{zeldov95,fuchs96,keener97} 
undergo a first order phase transition. 
This conclusion comes from latent heat measurements
\cite{schilling96} 
as well as jumps in the resistivity 
\cite{safar93,fuchs96,keener97} 
and in the magnetization 
\cite{liang96,welp96,zeldov95}. 
It has generally been assumed that this is a melting transition
from a vortex solid to a vortex liquid. 
In this paper we suggest the possibility that the
melting transition actually occurs in two steps as the temperature
increases; the first step is a first order premelting
transition from an ordinary vortex lattice to a soft solid
with a small but finite shear modulus, and the second step is the
first order melting of the soft solid into a vortex liquid.
In this paper we focus on the first step. 
We present an analytic theory of a first order
premelting transition in which the shear modulus
jumps discontinuously. The transition is induced
by interstitial and vacancy line defects in the 
vortex lattice which soften the shear modulus $c_{66}$. 
We find good agreement with the experimental curve of transition
temperature versus field, latent heat and magnetization
jumps for YBCO and BSCCO. In the soft solid phase 
the superconducting phase coherence along the field 
is destroyed by the wandering of the defect lines
which become entangled in the vortices of the soft solid lattice
\cite{frey94,nelson91}. 
However, since wandering is energetically
costly, the superconducting correlation length along the c--axis is
long. 
Finally we speculate
about the relation between our proposed two-step transition
and the well known peak effect 
\cite{kwok96,kwok94}.

Let us describe our scenario for premelting. 
Our approach follows that of Granato who showed that interstitial atoms
soften the shear modulus of ordinary crystals and lead
to a first order transition 
\cite{granato92}.
We start with a 
vortex lattice in a clean layered superconductor with a magnetic
field $H$ applied perpendicular to the layers along the
c-axis. We consider the vortices
to be correlated stacks of pancake vortices. 
We will assume that the transition is induced
by topological defect lines, i.e., vacancies and interstitials.
In a Delaunay triangulation 
\cite{preparata85}
a vacancy or an interstitial in a triangular lattice
is topologically equivalent to a pair of bound dislocations 
\cite{ryu96}
as well as to a twisted bond defect 
\cite{kim96}.
High temperature decoration experiments 
\cite{kim96}
and Monte Carlo simulations 
\cite{ryu96} 
have found such
defects to be thermally excited. The introduction
of these defects softens the elastic moduli. Since the energy to introduce
interstitials and vacancies is proportional to the elastic moduli, softening
makes it easier to introduce more defects. The softening also increases
the vibrational entropy of the vortex lattice which leads to a
premelting transition. The transition is driven
by the increased vibrational entropy of the vortex lines
of the lattice, and not by
the entropy of the wandering of the defect lines. In fact
Frey, Nelson and Fisher 
\cite{frey94} 
showed that a phase transition
driven by the entropy of wandering flux lines occurs at a much
higher magnetic field than what is observed experimentally. In the vicinity
of the experimentally observed first order phase transition,
wandering in the transverse direction by more than a lattice spacing
is energetically quite costly and therefore rare. (The energy scale
is set by $\epsilon_{o}s$ 
\cite{review:blatter,review:brandt}.
Here $s$ is the interplane spacing and $\epsilon_{o}$, the energy
per unit length of a vortex, is given by 
$\epsilon_{o}=(\phi_{o}/4\pi\lambda_{ab})^2$
where $\phi_{o}$ is the flux quantum and $\lambda_{ab}$ is the penetration
depth for currents in the {\it ab} plane.) 

Experimentally the resistivity jumps up at the transition from zero to
a finite value as the temperature increases. This is consistent with
our model since 
the soft solid will have a finite resistivity
due to the motion of interstitial (and vacancy) lines. The barrier
for the motion of interstitials is very small 
\cite{frey94} 
and is of order $10^{-3} E_{o}$ per unit 
length, where $E_{o}=2\epsilon_{o}$. 
The defect lines act like
a liquid of lines existing in a soft solid host. Notice that if
one tries to measure the shear modulus of such a system using
resistivity measurements, only the defect lines would move relative
to the pinned soft solid and one would deduce the shear modulus was zero
\cite{pastoriza95,wu97,kwok96}.

The first order transition is nucleated in a small
region by a local rearrangement of existing line segments. Slightly above the
premelting temperature $T_{p}$ a vortex line can distort and make an
interstitial and a vacancy line segment that locally create a soft solid. 
This is the
analog of a liquid droplet which nucleates melting of a crystal.
The role of the surface tension is played by the energy to connect
the interstitial segment to the rest of the vortex line.
This connection can be a Josephson vortex lying between planes or a
series of small pancake vortex displacements spread over several layers.
When the length $\ell$ of the interstitial and vacancy segments equals
the critical length $\ell_{c}$, the energy gained
by premelting equals the energy cost of the connections. 
When $\ell>\ell_{c}$, it is energetically favorable for the defect
segments grow to the length of the system.

To study premelting we assume that we have a vortex lattice with
interstitial and vacancy lines extending the length of the lattice.
Our goal is to find the free energy density as a function of the
concentration $n$ of defect lines.
The free energy density is $f=f_{o}+f_{w}+f_{vib}+f_{wan}$ where 
$f_{o}$ is the free energy density of a perfect lattice,
$f_{w}$ is the work needed to introduce a straight interstitial 
or vacancy line into the lattice,
$f_{vib}$ is the vibrational free energy density of the system, and
$f_{wan}$ is the free energy due to the wandering
of the defect lines over distances large compared to the lattice spacing.
We now examine these terms in detail.

$f_{o}$, the free energy density of a perfect rigid 
flux lattice, is given by the London term \cite{frey94,tinkham96}:
\begin{equation}
f_{o}=\frac{B^{2}}{8\pi}+\frac{B\phi_{o}}{32\pi^{2}\lambda_{ab}^{2}}
\ln\left(\frac{\eta\phi_{o}}{2\pi\xi_{ab}^{2}B}\right),\;\;
\frac{\phi_{o}}{4\pi\lambda_{ab}^{2}}\ll B\ll H_{c2}
\label{eq:london}
\end{equation}
where 
$B$ is the spatially averaged magnetic induction, 
$\xi_{ab}$ is the
coherence length in the $ab$ plane, and $\eta$ is 0.130519 for
a hexagonal lattice and 0.133311 for a square lattice \cite{frey94}.
For $B$ near $H_{c2}$, $f_{o}$ is given by
the Abrikosov free energy \cite{degennes89}
\begin{equation}
f_{o}=\frac{B^{2}}{8\pi}-\frac{(H_{c2}-B)^{2}}
{8\pi[1+(2\kappa^{2}-1)\beta_{A}]}
\label{eq:abrikosov} 
\end{equation}
where the Ginzburg-Landau parameter $\kappa=\lambda_{ab}/\xi_{ab}$,
and the Abrikosov parameter $\beta_{A}$ is 1.16 for a triangular lattice
and 1.18 for a square lattice.

To calculate $f_{vib}$, we follow Bulaevskii {\it et al.} \cite{bulaevskii92}. 
We denote the displacement
of the $\nu$th vortex pancake in the $n$th plane from its
equilibrium position by ${\bf u}(n,\bf{r}_{\nu})$ where
${\bf u}=(u_{x},u_{y})$ and the pancake position 
${\bf r}=(r_{x},r_{y})$. The Fourier transform
${\bf u}({\bf k},q)=\sum_{n \nu}\bf{u}(n,\bf{r}_{\nu})\exp[i(\bf{k}\cdot
\bf{r}_{\nu} + qn)]$. $\bf{k}=(k_{x},k_{y})$ and $q$ is the wavevector
along the c-axis. $f_{vib}=-(k_{B}T/V)\ln Z_{vib}$ where $V$ is the
volume and the vibrational partition function $Z_{vib}$ is given by
\begin{equation}
\ln Z_{vib} = \sum_{\bf{k},q>0,i}\ln\int\frac
{du_{R}(i {\bf k} q)du_{I}(i {\bf k} q)}
{\pi\xi_{ab}^2} e^{-{\cal F}_{el}/k_{B}T}
\label{eq:Z}
\end{equation}
where we have divided by the area $\pi\xi_{ab}^2$ of the normal
core of a pancake \cite{bulaevskii92}. $u_{R}$ and $u_{I}$ are the real
and imaginary parts of $\bf{u}(\bf{k},q)$ and $i\epsilon \{x,y\}$.
The elastic free energy functional associated with these distortions is 
\begin{equation}
{\cal F}_{el} =\frac{1}{2}\upsilon_{o}\sum_{{\bf k}q}
\sum_{ij}u_{i}(q,{\bf k})a_{ij}u_{j}^{*}(q,{\bf k})
\label{eq:elastic}
\end{equation}
where $i$ and $j\epsilon \{x,y\}$, the volume per
pancake vortex is $\upsilon_{o}=s\phi_{o}/B$, and
$s$ is the interplane spacing. 
The $\bf{k}$ sum is over a circular Brillouin
zone $K_{o}^{2}=4\pi B/\phi_{o}$. The matrix $a_{ij}$ is given by
$a_{ij}=c_{B}k_{i}k_{j}+(c_{66}k^{2}+c_{44}Q^{2})\delta_{ij}$
where $c_{B}$, $c_{66}$, and $c_{44}$ are the bulk, shear, and tilt
moduli, respectively. $c_{B}=c_{11}-c_{66}$ for a hexagonal lattice.
$Q^{2}=2(1-\cos qs)/s^{2}$. 
Diagonalizing $a_{ij}$ leads to 2 eigenvalues:
$A_{\ell}(kq)=c_{11}k^{2}+c_{44}Q^{2}$ and $A_{t}=c_{66}k^{2}+c_{44}Q^{2}$,
where $A$ is the diagonal matrix, the subscript $\ell$ 
denotes longitudinal and $t$ denotes transverse. Using this,
we can integrate over $u$ in (\ref{eq:Z});
the remaining sums over $\bf{k}$ and $q$ are done numerically.
At low fields ($b=B/H_{c2}<0.25$), the
elastic moduli are given by \cite{review:blatter,review:brandt}
\begin{eqnarray}
c_{66} & = & \frac{B\phi_{o}\zeta}{(8\pi\lambda_{ab})^{2}} \nonumber \\
c_{11} & = & \frac{B^{2}[1+\lambda_{c}^{2}(k^{2}+Q^{2})]}
{4\pi[1+\lambda_{ab}^{2}(k^{2}+Q^{2})](1+\lambda_{c}^{2}k^{2}
+\lambda_{ab}^{2}Q^{2})} \nonumber \\
c_{44} & = & \frac{B^{2}}{4\pi(1+\lambda_{c}^{2}k^{2}+\lambda_{ab}^{2}Q^{2})}
+ \frac{B\phi_{o}}{32\pi^{2}\lambda_{c}^{2}} \label{eq:moduli} \\
       &   & \mbox{} \times \ln\frac{\xi_{ab}^{-2}}{K_{o}^{2}+(Q/\gamma)^{2}
+\lambda_{c}^{-2}} + \frac{B\phi_{o}}{32\pi^{2}\lambda_{ab}^{4}Q^{2}}
\ln(1+\frac{Q^{2}}{K_{o}^{2}}) \nonumber \\
\nonumber
\end{eqnarray}
where $\lambda_{c}$ is the penetration depth for currents along the
c-axis, $\gamma=\lambda_{c}/\lambda_{ab}$ is the anisotropy, and $\zeta=1$. 
At high fields ($b>0.5$) \cite{review:blatter,review:brandt}, 
$c_{66}$ is altered by the factor
$\zeta\approx(1-0.5\kappa^{-2})(1-b)^{2}(1-0.58b+0.29b^{2})$
and the penetration depths in $c_{11}$ and $c_{44}$ are replaced by
$\tilde{\lambda}^{2}=\lambda^{2}/(1-b)$ 
where $\lambda$ denotes either $\lambda_{ab}$ or $\lambda_{c}$.
In addition the last two terms of $c_{44}$ are replaced by
$B\phi_{o}/(16\pi^{2}\tilde{\lambda}_{c}^{2})$. These
replacements guarantee that the elastic moduli vanish at $H_{c2}$.
For YBCO
the temperature dependence of the penetration depths and coherence
lengths are given by $\lambda(T)=\lambda(0)(1-(T/T_{c}))^{-1/3}$
\cite{kamal94} and $\xi_{ab}(T)=\xi_{ab}(0)(1-(T/T_{c}))^{-1/2}$, 
respectively.
For BSCCO whose premelting field is two orders 
of magnitude below $H_{c2}$, 
$\lambda^{2}(T)=\lambda^{2}(0)/(1-(T/T_{c})^{4})$ and
$\xi_{ab}^{2}(T)=\xi_{ab}^{2}(0)/(1-(T/T_{c})^{4})$ \cite{tinkham96}.

The free energy density $f_{w}$ due to the energy cost
of adding a vacancy or interstitial vortex line is difficult
to calculate accurately \cite{frey94}. However, we can
write down a plausible form for $f_{w}$ by noting that a straight
line defect parallel to the c-axis
produces both shear and bulk (but not tilt) distortions
of the vortex lattice. For example, if a defect at the origin produces
a displacement $\bf{u}$ that satisfies 
$\nabla\cdot{\bf u}=\upsilon_{o}\delta({\bf r})/s$ where
$\delta({\bf r})$ is a two dimensional delta function, then
$u_{\alpha}({\bf k})=ik_{\alpha}/k^{2}$ \cite{frey94}. Inserting
this in (\ref{eq:elastic}), we find that 
$f_{w}=(c_{66}+\overline{c}_{B})/2$ where 
$\overline{c}_{B}=\sum_{\bf{k}}c_{B}(q=0,\bf{k})$. Generalizing
this to allow for a more complicated distortion
and for a concentration $n$ of line defects, we write \cite{granato92}
\begin{equation}
f_{w}=\int_{0}^{n} dn(\alpha_{1}c_{66}+\alpha_{2}\overline{c}_{B})
\label{eq:fw}
\end{equation}
where $\alpha_{1}$ and $\alpha_{2}$ are dimensionless constants.
We expect the isotropic distortion to be small, i.e., $\alpha_{2}\ll 1$,
and the shear deformation to dominate, i.e., $\alpha_{1}\gg\alpha_{2}$.
Integrating over $n$ allows the elastic moduli to depend
on defect concentration. We will assume that $c_{B}$ is independent of
$n$ since we believe that the bulk modulus of the vortex solid is
roughly the same as that of the soft solid phase. To find $c_{66}(n)$
\cite{granato92}, we use its definition 
$c_{66}=\partial^{2}f/\partial\varepsilon^{2}$ where
$\varepsilon$ is the shear strain. Assuming
that $c_{B}$ has negligible shear strain dependence, we find
$c_{66}(n)=c_{66}(0)+\alpha_{1}\int_{0}^{n}(\partial^{2}c_{66}(n)/
\partial\varepsilon^{2})dn$ or
\begin{equation}
\frac{\partial c_{66}(n)}{\partial n} = \alpha_{1}\frac{\partial^{2}c_{66}(n)}
{\partial\varepsilon^{2}}
\label{eq:c66}
\end{equation}
If we shear the lattice in the $ab$ plane along rows separated by a distance
$d$, the shear modulus must be periodic in displacements equal to
the lattice constant $a_{o}$. We describe this with the simplest
even periodic function: $c_{66}(u)=c_{66}(u=0)\cos(2\pi u/a_{o})=
c_{66}(\varepsilon=0)\cos(2\pi d\varepsilon/a_{o})$ where $\varepsilon=u/d$.
Then $\partial^{2}c_{66}(n)/\partial\varepsilon^{2}=-\beta c_{66}(n)$,
where $\beta=4\pi^{2}d^{2}/a_{o}^{2}$. Combining this with
(\ref{eq:c66}), we obtain $c_{66}(n)=c_{66}(0)\exp(-\alpha_{1}\beta n)$.
Thus the shear modulus softens exponentially with the defect
concentration $n$. This softening lowers the energy cost to introduce
further defects, and increases the vibrational free energy $f_{vib}$
when $c_{66}(n)$ is used in $a_{ij}$.
Substituting $c_{66}(n)$ into our expression (\ref{eq:fw}) for
$f_{w}$ yields
\begin{equation}
f_{w}=\frac{c_{66}(n=0)}{\beta}(1-e^{-\alpha_{1}\beta n}) +
\alpha_{2}\overline{c}_{B}n
\label{eq:finalfw}
\end{equation}

The last term we need to consider is $f_{wan}$, the free energy 
due to the wandering
of the defect lines over distances large compared to the lattice spacing.
We can estimate $f_{wan}$ with the following expression \cite{frey94}
\begin{equation}
f_{wan}\approx -\frac{k_{B}T}{\ell_{z}a_{o}^{2}}\ln(m_{\ell})
\label{eq:wandering}
\end{equation}
where $m_{\ell}=3$ for a triangular lattice (BSCCO) and $m_{\ell}=4$
for a square lattice (YBCO). $\ell_{z}$ can be thought of as the
distance along the z--axis that it takes for the defect line to wander
a transverse distance of one lattice spacing $a_{o}$. To go from
one vacancy or interstitial site to the next, the defect line segment
must jump over the barrier between the two positions. This gives
$\ell_{z}$ a thermally activated form: 
$\ell_{z}\sim \ell_{o}\exp(-E/k_{B}T)$, where 
$\ell_{o}\approx a_{o}(\epsilon_{1}/\epsilon_{B})^{1/2}$ and
$E\approx a_{o}(\epsilon_{1}\epsilon_{B})^{1/2}$.
$\epsilon_{1}$ is the line tension and is given by
$\epsilon_{1}\sim (\epsilon_{o}/\gamma^{2})\ln(a_{o}/\xi_{ab})$.
Numerical simulations \cite{frey94} indicate that
the barrier height $\epsilon_{B}$ is small and we use
$\epsilon_{B}=2.5\cdot 10^{-3}\epsilon_{o}$.
$f_{wan}$ itself is quite small compared to the other terms because of
the high energy cost of vortex displacements. For example,
in the soft solid phase at the transition $f_{wan}$ is about
two orders of magnitude smaller than $f_{w}$ or $f_{vib}$.
Thus the transition is not
driven by a proliferation of wandering defect lines because
near the transition the high energy cost of vortex displacements is not
sufficiently offset by the entropy of the meandering line \cite{frey94}.

Before we plot $f$ versus $n$, we note that the difference
between $B$ and $H$ is negligible for YBCO but can be a significant
fraction of the premelting field $H_{p}$ for BSCCO. To find the
value of $B$ to use in the Helmholtz free energy density $f$,
we minimize the Gibbs free energy density $G$, i.e.,
$\partial G/\partial B=0$ where $G=f-{\bf B}\cdot{\bf H}/4\pi$.
Since the concentration dependence of $B$ is negligible, we find
$B$ for $n=0$ for each value of $H$ and $T$. Typical plots of
$\Delta f=f(n)-f(0)=f_{w}+\Delta f_{vib}$ versus $n$ 
are shown in the inset of figure 1. The
double well structure of $\Delta f$ is characteristic of a first
order phase transition. The equilibrium transition occurs 
when both minima have the same
value of $\Delta f$. We associate the minimum at $n=0$
with the vortex solid and the minimum at finite $n$ with a soft vortex
solid that has a small but finite shear modulus. 
The defect concentration at the transition
is only a few percent. At higher concentrations
$\Delta f$ increases with increasing $n$ because
introducing defects costs compressional energy which
is proportional to the bulk modulus. Thus defects
do not proliferate. As an estimate of the softness
at the transition, for
$n=5$\%, $c_{66}(n)\sim 0.2c_{66}(0)$ for BSCCO.
The strain field $\varepsilon_{\alpha\beta}^{\rm d}(k)$
produced by the defect determines whether the
shear modulus is zero in the high temperature
phase \cite{marchetti90}. For dislocation
loops, $\varepsilon_{\alpha\beta}^{\rm d}(k)$ is singular as
$k\rightarrow 0$, and the shear modulus is zero at $k=0$
\cite{marchetti90}. For vacancy and interstitial lines 
$\varepsilon_{\alpha\beta}^{\rm d}(k)$ is finite, and hence the shear
modulus is nonzero.

In Figure \ref{fig:melting} we fit
the experimental first order transition curves in the $H-T$ plane
using 2 adustable parameters: $\alpha_{1}$ and $\alpha_{2}$.
As expected, $\alpha_{1}\gg\alpha_{2}$ and $\alpha_{2}\ll 1$
(see Figure \ref{fig:melting}). The geometrical 
quantity $\beta$ can have several
values for a given lattice structure, depending on which planes
are sheared. We choose $\beta=\pi^{2}\tan^{2}\phi$ where
$\phi$ is the angle between primitive vectors.
Decoration experiments on BSCCO find a triangular
lattice \cite{kim96}, so we use $\phi=60^{o}$. 
For YBCO we choose $\phi=44.1^{o}$ which is very close
to a square lattice which has $\phi=45^{o}$. Maki \cite{won96}
has argued that the $d$-wave symmetry of the order parameter
yields a square vortex lattice tilted by 45$^{o}$ from the
$a-$axis. Experiments \cite{yethiraj93,keimer94,maggio-aprile95} 
on YBCO find $\phi$ ranging from  
$36^{o}$ to $45^{o}$. 

We can calculate the jump in magnetization 
$\Delta M$ at the transition using
$\Delta M = -\partial\Delta G/\partial H|_{T=T_{p}}$. The jump
in entropy $\Delta s$ is given by 
$\Delta s=-\upsilon_{o}\partial\Delta G/\partial T|_{H=H_{p}}$ where 
$\Delta s$ is the entropy change per vortex per layer. The results
are shown in Figure \ref{fig:mag-S}. 
We have checked that our results
satisfy the Clausius-Clapeyron equation 
$\Delta s=-(\upsilon_{o}\Delta B/4\pi)dH_{p}/dT$. We obtain
good agreement with experiment well below $T_{c}$. Near $T_{c}$
it is thought that the entropy jump is enhanced by microscopic
degrees of freedom \cite{dodgson97,rae97} which are not included
in our model.

We can compare our results with the Lindemann criterion by
calculating the mean square displacement $<|u|^{2}>$ at the transition
using eq. (\ref{eq:Z}):
$<|u|^{2}>=-(2k_{B}T/\upsilon_{o})\sum_{\alpha{\bf k}q}
\partial\ln Z_{vib}/\partial A(\alpha{\bf k}q)$ where
$A$ is the diagonal matrix similar to $a_{ij}$ and $\alpha$
labels the 2 eigenvalues. Defining the Lindemann ratio $c_{L}$
by $c_{L}^{2}=<|u|^{2}>/a_{o}^{2}$, we find that $c_{L}\approx 0.25$
for YBCO at $H_{p}=5$ T and that $c_{L}\approx 0.11$ for BSCCO at
$H_{p}=200$ G. 
Here we have used the same values of the parameters
that were used to fit the phase transition
curves in Figure \ref{fig:melting}. These
values of $c_{L}$ are consistent with previous values 
\cite{review:blatter,review:brandt,houghton89}.

Experiments have found little, if any, hysteresis 
\cite{safar93,zeldov95,keener97}. This is consistent with
our calculations. We can bound 
the hysteresis by noting the range of temperatures between
which the soft solid minimum appears and the solid minimum disappears.
Typical values for the width of this temperature 
range are 300 mK for YBCO at $H=5T$ and 1.3 K for
BSCCO at $H=200$ G. Another measure of the hysteresis can be found
in the plots of $\Delta f$ versus $n$.
The barrier height $V_{B}$ between the minima
is low ($V_{B}\upsilon_{o}\sim 30$ mK) which is consistent with minimal
hysteresis. 

In going from the normal metallic phase to the vortex solid, two
symmetries are broken: translational invariance and gauge
symmetry which produces the superconducting
phase coherence along the magnetic field.
In the soft solid phase, longitudinal superconductivity is
destroyed by the wandering of the defect lines
which become entangled with the soft solid vortices.
(A vortex solid with entangled vortex lines has been termed
a supersolid \cite{frey94,nelson91}.)
Even though line wandering is energetically costly
and therefore rare, it does occur. As a result, the correlation
length along the c--axis will be quite long. 
This is consistent with measurements
in YBCO of the c--axis resistivity which find that there is loss of
vortex velocity correlations
for samples thicker than 100 $\mu$m \cite{lopez96,lopez96a,lopez96b}.
For samples thicker than the longitudinal correlation length, 
the loss of longitudinal
superconductivity coincides with the premelting transition \cite{chen95}. 
This agrees with experiments which indicate that the
loss of superconducting phase coherence along the c--axis
coincides with the first order transition \cite{lopez96,lopez96a,lopez96b}.

Since the soft solid is a lattice with a few percent of defect lines,
the Fourier transform of the density-density correlation function
should exhibit Bragg peaks. Relative to the ordinary
vortex solid, the intensity of these
peaks would be slightly diminished by the defect lines, so it would
be difficult to detect the transition via neutron scattering.
In going from the soft solid to the normal metallic state,
translational invariance is regained by a first order melting
transition. Thus there are two transitions: the premelting
transition and the melting of the soft solid. Melting is
observable in small angle neutron scattering
experiments \cite{cubitt93} which see a rapid decrease in the intensity of
the Bragg spots. The region of the phase diagram where the soft solid
exists may be quite narrow, of order a degree or less in temperature
\cite{kwok94}.
There is the intriguing possibility that our scenario of two transitions
may be related to the peak effect in which the critical current as a
function of temperature or field is observed to sharply increase
below the melting transition \cite{kwok94}. This increase is believed
to result from the enhanced pinning of flux lines due to the softening of the
shear modulus $c_{66}$ \cite{larkin95}.

To summarize we have discussed the possibility that a vortex lattice
melts in two stages. First it 
undergoes a first order premelting transition into a soft solid
followed by another first order phase transition into a liquid.
The premelting transition
is induced by vacancy and interstitial vortex lines that soften
the shear modulus and enhance the vibrational entropy. 
The entanglement of these defect lines with
the vortex lines of the soft solid leads
to the loss of longitudinal superconducting phase coherence. 
However, the correlation length
corresponding to longitudinal superconductivity is quite long
because line wandering is energetically costly and therefore rare.
We obtain good agreement with the experimentally
measured curve of transition temperature versus field, latent heat, 
and jumps in magnetization
for BSCCO and YBCO. The Lindemann ratio $c_{L}$ is $\sim 11$\% for BSCCO
and $\sim 25$\% for YBCO. The hysteresis
is small.

We thank Andy Granato, Marty Maley, Lev Bulaevskii, Sue
Coppersmith, and David Nelson for helpful
discussions. This work was supported in part by ONR grant
N00014-96-1-0905 and by funds provided by the University of
California for the conduct of discretionary research by Los
Alamos National Laboratory. HMC acknowledges support by the Swiss 
Nationalfonds.


\begin{figure}
\caption{First order phase transition curves of 
magnetic field versus temperature.
for YBCO and BSCCO. Parameters used for YBCO are $\alpha_{1}=2.55$,
$\alpha_{2}=0.01485$, $\phi=44.1^{o}$, $\lambda_{ab}(0)=1186 \AA$ 
\protect\cite{kamal94},
$s=12 \AA$, $\xi_{ab}(0)=15 \AA$, $\gamma=5$, and $T_{C}=92.74$ K.
Parameters used for BSCCO are $\alpha_{1}=1.0$, $\alpha_{2}=0.00705$,
$\phi=60 ^{o}$, $\lambda_{ab}(0)=2000 \AA$, $s=14 \AA$, 
$\xi_{ab}(0)=30 \AA$,
$\gamma=200$, and $T_{C}=90$ K. For BSCCO we use the
low field form of the elastic moduli from (\protect\ref{eq:moduli})
and for YBCO we use the
high field form. For $f_{o}$ we use (\protect\ref{eq:london})
for BSCCO and (\protect\ref{eq:abrikosov}) for YBCO.
(For BSCCO we plot $B$ vs. $T$ because
that is what ref. \protect\cite{zeldov95} measured.) 
The experimental points
for YBCO come from ref. \protect\cite{schilling96} and those for BSCCO come
from ref. \protect\cite{zeldov95}. Inset: Typical $\Delta f$ versus $n$.}
\label{fig:melting} 

\caption{(a) and (b): Entropy jump $\Delta s$ per vortex per layer
versus $T_{p}$ at the transition
for YBCO and BSCCO. The experimental points for YBCO are from
\protect\cite{schilling96} and
those for BSCCO are from \protect\cite{zeldov95}.
(c) and (d): Magnetization jump $\Delta M$ versus $T_{p}$ at the first
order phase transition for YBCO and BSCCO. 
The experimental points for YBCO are from \protect\cite{welp96} and 
those for BSCCO are from \protect\cite{zeldov95}. For the
theoretical points the values of the 
parameters are the same as in Figure 1 for all the curves.}
\label{fig:mag-S}

\end{figure} 

\end{multicols}
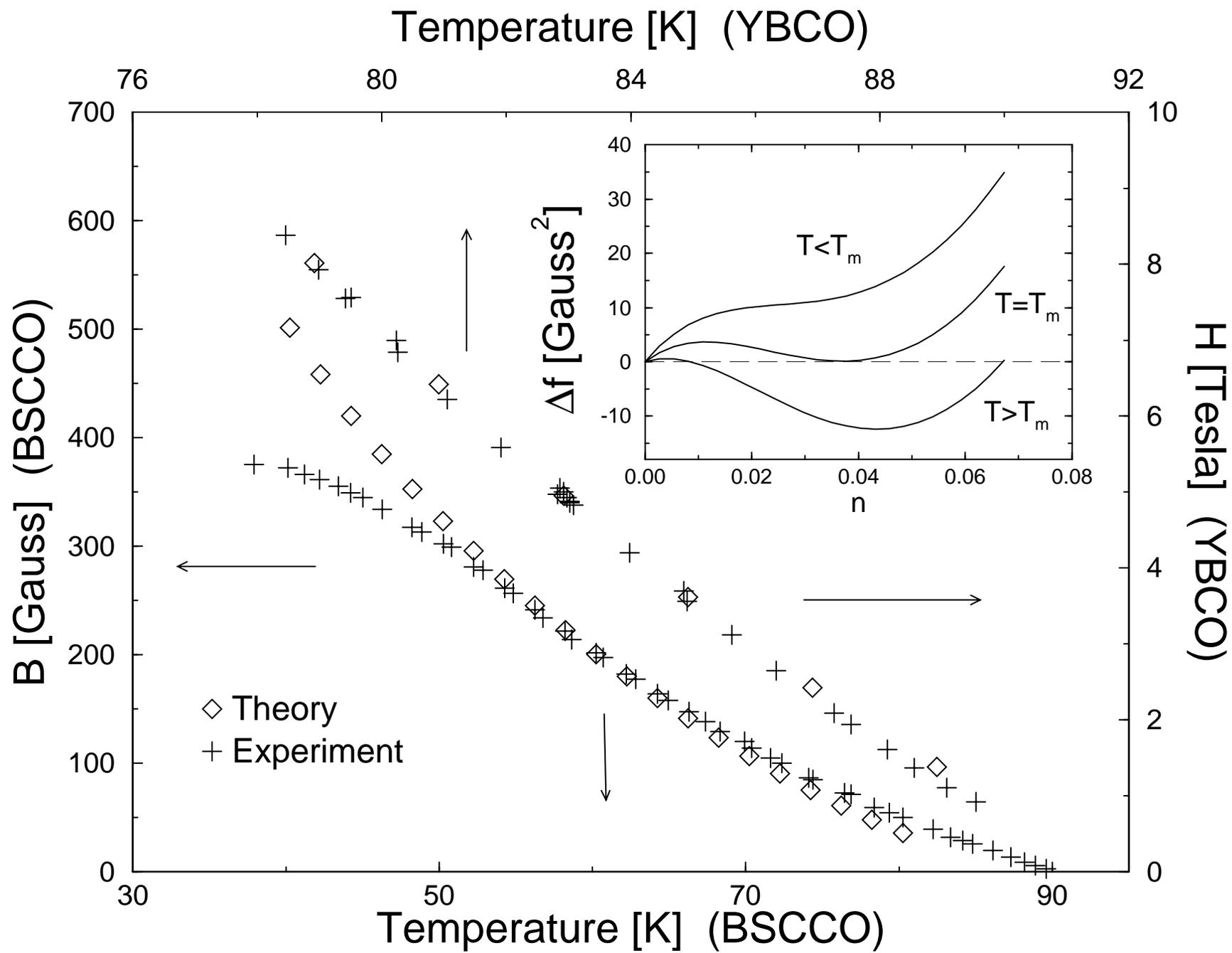
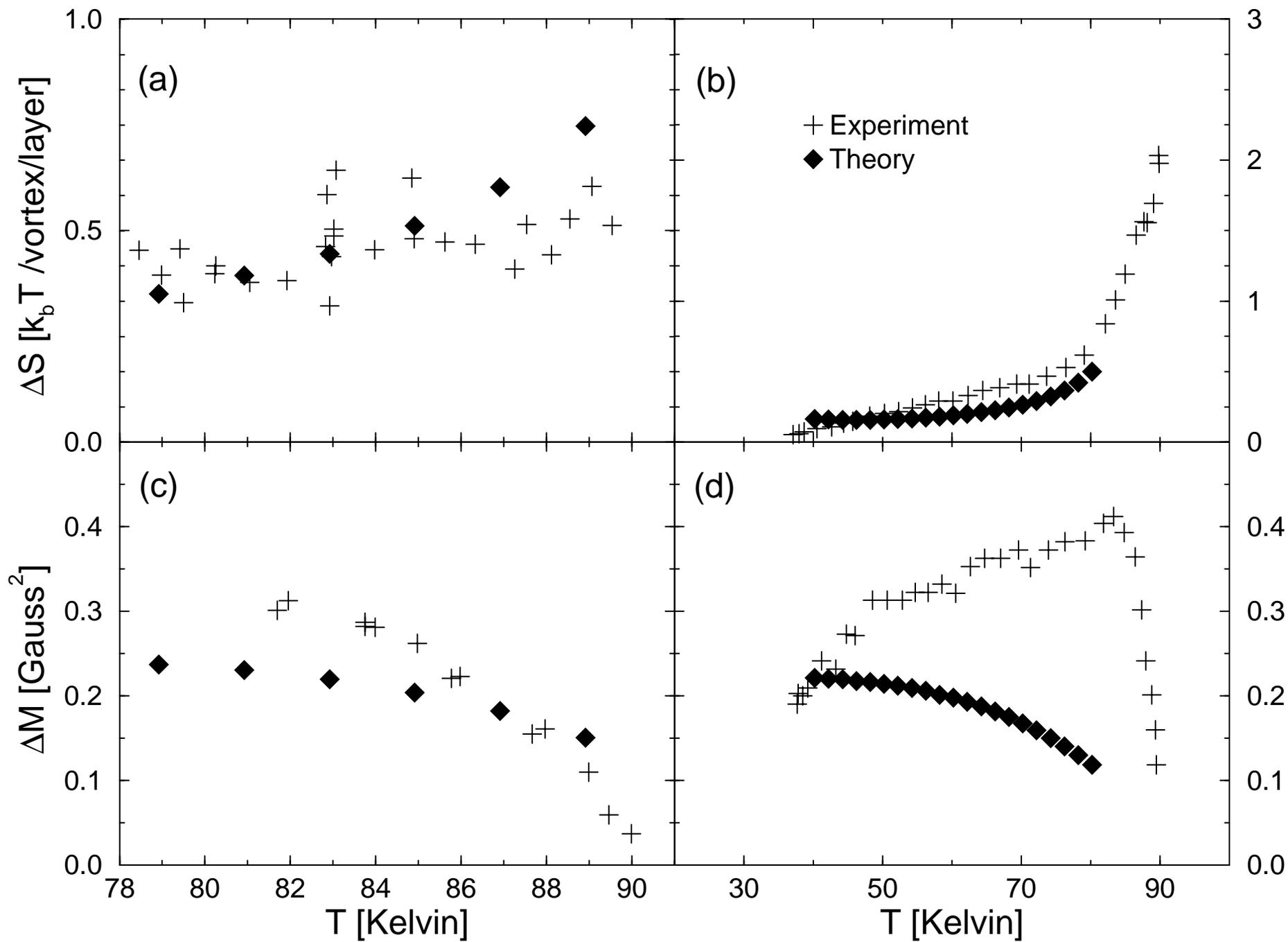
\end{document}